# Mathematical approaches to differentiation and gene regulation

*Christophe Soulé*

*Institut des Hautes Études Scientifiques*
*35, route de Chartres*
*F - 91440 Bures-sur-Yvette*

Email : soulé@ihes.fr

**Abstract**

We consider some mathematical issues raised by the modelling of gene networks. The expression of genes is governed by a complex set of regulations, which is often described symbolically by interaction graphs. These are finite oriented graphs where vertices are the genes involved in the biological system of interest and arrows describe their interactions: a positive ( resp. negative) arrow from a gene to another represents an activation (resp. inhibition) of the expression of the latter gene by some product of the former. Once such an interaction graph has been established, there remains the difficult task to decide which dynamical properties of the gene network can be inferred from it, in the absence of precise quantitative data about their regulation. There mathematical tools, among others, can be of some help.

In this paper we discuss a rule proposed by R.Thomas according to which the possibility for the network to have several stationary states implies the existence of a positive circuit in the corresponding interaction graph. We prove that, when properly formulated in rigorous terms, this rule becomes a theorem valid for several different types of formal models of gene networks. This result is already known for models of differential [8] or boolean [9] type. We show here that a stronger version of it holds in the differential setup when the decay of protein concentrations is taken into account. This allows us to verify also the validity of Thomas' rule in the context of piecewise-linear models. We then discuss open problems.

**Keywords** : Gene network, regulation, differentiation, interaction graph, positive circuit, model, boolean network, differential system, piecewise-linear differential inclusion.

In living organisms, many proteins are transcription factors: they can bind to DNA and regulate the transcription of specific genes, i.e. the synthesis of RNA from coding regions of chromosomal DNA. This regulation of transcription is a very complex mechanism, which can involve up to dozen of genes, and other factors as well. Furthermore, regulation occurs during the full process of gene expression. In all cases, one will say that a gene *A* activates (resp. inhibits) a gene *B* when *A* produces a protein which has a positive (resp. negative) effect on the expression of gene *B*. If several genes are involved in a given biological system, they form a gene network from which one can draw an *interaction graph G*. In mathematical terms, *G* is a finite oriented graph, the edges of which are endowed with a sign: the vertices are the genes, and a positive (resp. negative) edge from *j* to *i* means that *j* activates (resp. inhibits) *i*. Note that a gene can activate or inhibit itself, i.e. *G* can have edges which end where they start. Furthermore, depending on the concentrations of the proteins in the system, the effect of *j* on *i* can be positive, negative, or absent. In other words *G* is a function of the concentrations.

In general, very little is known about the strength of the interactions between genes. One is thus faced with the following difficult problem: which dynamical properties of a gene network can be inferred from the topology of its interaction graph (despite the lack of quantitative information). Several methods have been used to tackle this question. One of them is numerical simulation, which requires to choose kinetic parameters in a realistic way. Another method is to study the statistical properties of gene networks, by comparing their interaction graphs with random ones. One can also try to decompose a given graph into submodules of biological significance. Finally, some authors have focused their attention on special motifs, i.e. subgraphs with simple topology, for instance those involving few vertices which are overrepresented in gene networks [1].

In this paper we shall study *circuits*. A circuit *C* in an interaction graph *G* is a sequence $e_1,\ldots, e_k$ of edges such that the end point of $e_i$, $i = 1,\ldots, k - 1$ (resp. $e_k$) is the origin of $e_{i+1}$ (resp. $e_1$), each vertex of *G* occuring at most once in *C*. The sign of a circuit is the product of the signs of its edges. When gene regulation was discovered, it was soon noticed that circuits (or 'feedback loops') are often present in gene networks (or at least in mixed networks, with interactions between genes, proteins and metabolites), and that their biological role depends on their sign. For instance, if *G* = *C* consists of a single positive circuit, the network can have two possible stable stationary states. For instance, when *G* looks as follows

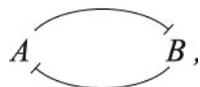

where ⊣ denotes a negative edge, common sense suggests that either *A* or *B* will win the competition and *B* (resp. *A*) will be shut off. Now, according to an idea of Delbrück [2], the possibility for a gene network to have several stationary states is one possible mechanism for biological differentiation.

However, when the interaction graph *G* contains a positive circuit *C*, there is no reason to expect that *C* will govern the dynamic behaviour of the underlying network; this depends on the strength of the interactions. R. Thomas [3] had the idea to turn the statement another way: positive circuits are necessary if not sufficient for multistationarity. He proposed the following:

*Thomas' rule: Assume a gene network has several nondegenerate stationary states. Then its interaction graph contains, somewhere in phase space, a positive circuit.*

Here, the expression 'somewhere in phase space' refers to the fact, mentioned earlier, that the concentrations of proteins have to be given for the graph to be defined.

This rule of Thomas, if true, can be quite useful for geneticists. For instance, if one knows that a given biological system can differentiate, one will be led to search for a positive circuit relating the genes involved. It is therefore worthwhile to decide how general this rule is. One way to check it is the following. First, describe a gene network in a formal mathematical way (a model). Then, within this formalism, make sense of all the terms figuring in Thomas' rule (nondegenerate stationary state, interaction graph, phase space…). Thomas' rule then becomes, in this context, a precise mathematical statement (a conjecture), which one can try to prove (or to contradict) logically.

For instance, R. Thomas and M. Kaufman phrased Thomas' rule [3] as a precise mathematical conjecture by using a differential model of gene networks [4]. This conjecture was proved under additional assumptions in [5] [6] [7], and in general in [8]. Later, Thomas' rule was checked for boolean models [9]. Many more models of gene networks have been proposed (see [10] for a thorough survey). In this paper, we shall show that Thomas' rule is true for five different types of models of gene networks: the boolean, differential, differential with decay, piecewise-linear and multivalued discrete models. Of course, knowing that the rule is true for one type of models does not imply automatically that it is true for another one. Still, our arguments will exhibit interesting connections between several ways of modelling. When the spontaneous decay of all proteins is taken into account, Thomas' rule happens to be more robust, and this allows us to study the piecewise-linear case by approximation. And, in some cases, the piecewise-linear models can in turn be described in discrete terms. We conclude the paper by discussing open problems. Among them is whether Thomas' rule remains valid for other biological networks, or when the stochastic nature of gene regulation is taken into account.

**Acknowledgements:** I wish to thank H. De Jong, J. Demongeot, M. Kaufman, N. Le Novère, O. Radulescu, E. Rémy, P. Ruet, D. Thieffry, R. Thomas and A.Wagner for helpful discussions and comments .

## 1. Boolean models

Let $n \geq 1$ be an integer and $\Omega = \{0,1\}^n$, the set of strings of $n$ letters in the alphabet $\{0,1\}$. Consider a map

$$F = (F_i) : \Omega \to \Omega .$$

The pair $(\Omega, F)$ is usually called a boolean network.

[According to S. Kauffman [11], we can view the data $(\Omega, F)$ as a model for the dynamic of a network of $n$ genes. A point $x = (x_i) \in \Omega$ describes a state of the network: $x_i = 1$ (resp. $x_i = 0$) when the gene $i$ is active (resp. inactive). The map $F$ describes the evolution of this network: if it is in state $x$ at a given time $t$, it will be in state $F(x)$ at time $t + 1$].

To every $x$ in $\Omega$ we attach an interaction graph $G(x)$ which is described as follows. Fix $j \in \{1,...,n\}$ and let $y \in \Omega$ be defined by

$$y_j = 1 - x_j \quad \text{and} \quad y_k = x_k \quad \text{if} \quad k \neq j .$$

Given $i \in \{1,...,n\}$, there is an edge from $j$ to $i$ when $F(y)_i \neq F(x)_i$. This edge is positive if $x_j = F(x)_i$ and it is negative otherwise. [To illustrate this definition, assume that $x_j = 1$, i.e. gene $j$ is active in $x$ and inactive in $y$. If $F(x)_i = x_j = 1$ and $F(y)_i = 0$, we can say that, by inhibiting $j$ in $x$ we have inhibited $i$ in $F(x)$. In other words $j$ is an activator of $i$ in the state $x$, and we have a positive edge from $j$ to $i$ in the graph $G(x)$].

A stationary state of the network is a fixed point of $F$, i.e. a point $x \in \Omega$ such that $F(x) = x$. Part 2) of the following theorem says that Thomas' rule is true for boolean models:

*Theorem 1.*
1) [10] Assume that none of the graphs $G(x)$, $x \in \Omega$, contains a circuit. Then $F$ has a unique fixed point.
2) [12] Assume that $F$ has several fixed points. Then there exists $x \in \Omega$ such that $G(x)$ contains a positive circuit.

## 2. Differential models

Let $n \geq 1$ be an integer and $\Omega = \mathbf{R}^n$ the standard real vector space of dimension $n$. Consider a differential map

$$F = (F_i) : \Omega \to \Omega ,$$

and the system of differential equations

(1)
$$\frac{dx}{dt} = F(x),$$

where $x : \mathbf{R} \to \mathbf{R}^n$ is any differential path in $\Omega$.

[According to R. Thomas and M. Kaufman [4], this is a model for a network of $n$ genes. For every $i = 1, \ldots, n$, the number $x_i(t)$ is the concentration of the protein $i$ at time $t$. The equation (1) says that the variation of $x_i(t)$ is a function of all the concentrations $x_j(t)$, $j = 1, \ldots, n$].

Given $x \in \Omega$, we define an interaction graph $G(x)$ as follows. Its set of vertices is $\{1,\ldots,n\}$ and there is a positive (resp. negative) edge from $j$ to $i$ when the partial derivative $\frac{\partial F_i}{\partial x_j}(x)$ is positive (resp. negative). [To illustrate this, assume $\frac{\partial F_i}{\partial x_j}(x) > 0$. If we increase the concentration of protein $j$ in state $x$, the number $F_i(x)$ will increase, and the production of $i$ will accelerate. In other words, $j$ is an activation of $i$ in state $x$].

From (1) we see that a stationary state of the network is a zero of $F$, i.e. $x \in \Omega$ such that $F(x) = 0$. This zero is called nondegenerate when the determinant $\det\left(\frac{\partial F_i}{\partial x_j}(x)\right)$ of the Jacobian matrix at $x$ is different from zero. Thomas' rule is true for differentiable models :

***Theorem 2.*** [8] Assume that $F$ has at least two nondegenerate zeroes. Then there exists $x \in \Omega$ such that $G(x)$ contains a positive circuit.

For previous results see [5] [6] [7].

### 3. Differentiable models with decay

Since concentrations cannot be negative, one would like to get a version of Theorem 2 where $F(x)$ is only defined for those $x = (x_i)$ such that $x_i \geq 0$, $i = 1, \ldots, n$. But it turns out that it is not true as stated with this restriction ([8], 3.5). However, a more realistic modelling of gene networks consists in taking into account that the concentration of every protein is submitted to a spontaneous decay, due to degradation and to the growth of cells. We are thus led to the following model.

For every $i = 1, \ldots, n$ let $\Omega_i \subset \mathbf{R}$ be a real interval (i.e. $\Omega_i = [a_i, b_i]$, $]a_i, b_i]$, $[a_i, b_i[$ or $]a_i, b_i[$, with $-\infty \leq a_i$ and $b_i \leq +\infty$). On the product $\Omega = \prod_i \Omega_i \subset \mathbf{R}^n$ consider a differentiable map

$$F = (F_i) : \Omega \to \mathbf{R}^n$$

and the system of differential equations

(2) $$\frac{dx_i}{dt} = F_i(x) - \gamma_i x_i, \quad i = 1, \ldots, n,$$

where $\gamma_1 > 0, \ldots, \gamma_n > 0$ are fixed constants [the degradation rates].

For every $x \in \Omega$, let $G(x)$ be the interaction graph defined from the signs of the partial derivatives $\frac{\partial F_i}{\partial x_j}(x)$ as in §2 above.

**Theorem 3.** Assume that there exists two points $x \neq y$ in $\Omega$ such that

$$\left| F_i(x) - \gamma_i x_i - F_i(y) + \gamma_i y_i \right| < \gamma_i \left| x_i - y_i \right|$$

for all indices $i$ such that $x_i \neq y_i$. Then there exists a point $z$ in $\Omega$ such that $G(z)$ contains a positive circuit.

**Remarks.** 1) A stationary state is a point $x \in \Omega$ such that, for all $i = 1, \ldots, n$,

$$F_i(x) - \gamma_i x_i = 0.$$

If $x \neq y$ are two stationary states of $(x)$ we have, for all $i = 1, \ldots, n$,

$$F_i(x) - \gamma_i x_i = F_i(y) - \gamma_i y_i = 0,$$

therefore the hypotheses of Theorem 3 are satisfied. In other words, Thomas' rule is again true, for arbitrary $\Omega$, when we take the spontaneous decay of proteins into account. It is also more "robust", since it remains valid for two states $x \neq y$ which are almost stationary.

2) In the conclusion of Theorem 3, we can assume that $z_i = x_i$ whenever $x_i = y_i$. To see that, let $I$ be the set of indices $i$ such that $x_i = y_i$, and apply Theorem 3 to the restriction of the functions $F_i$, $i \notin I$, to the linear subspace of those $z \in \mathbf{R}^n$ such that $z_i = x_i$ when $i \in I$. A similar remark can be made in Theorems 1, 2 and 4.

**4. Piecewise-linear models**

Let $\Omega = \prod_i \Omega_i$ be as in §3. Another model for genes networks [13] [14] [15] is given by the system of equations

$$\frac{dx_i}{dt} = F_i(x) - \gamma_i x_i,$$

where $\gamma_i > 0$ and each function $F_i$ is is a polynomial combination of step functions.

More precisely, for every real number $\theta$, define the step function $s(x, \theta)$ from **R** to subsets of **R** by

$$s(x, \theta) = \begin{cases} 1 & \text{if } x > \theta \\ ]0,1[ & \text{if } x = \theta \\ 0 & \text{if } x < \theta \end{cases}$$

For every $j = 1, \ldots, n$, choose finitely many distinct real thresholds $\theta_j^k$ in the interior of $\Omega_j$, $k = 1, \ldots, m_j$. For each $i = 1, \ldots, n$, fix a real polynomial $P_i(T_j^k)$ in $\sum_{j=1}^{n} m_j$ variables and, for every $x$ in $\Omega$, let

(3) $$F_i(x) = P_i(s(x_j, \theta_j^k)).$$

By definition, $F_i(x)$ is a subset of **R**, reduced to a single point if $x_j \neq \theta_j^k$ for all $j$ and $k$.

Now we consider the system of piecewise linear differential inclusions

(4) $$\frac{dx_i}{dt} \in F_i(x) - \gamma_i x_i, \quad i = 1, \ldots, n,$$

for some fixed real constants $\gamma_1 > 0, \ldots, \gamma_n > 0$.

A stationary state of (4) is a point $x$ in $\Omega$ such that 0 lies in $(F_i(x) - \gamma_i x_i)$. For every $x$ in $\Omega$ we define an interaction graph $G(x)$ as follows. Its set of vertices is $\{1,\ldots,n\}$ and there is a positive (resp. negative) edge from $j$ to $i$ when $x_j = \theta_j^k$ is a threshold and the value of the partial derivative $\partial P_i / \partial T_j^k$ is positive (resp. negative) at the some point in the set $(s(x_a, \theta_a^b))$. Note that in the case considered in [16] the interaction graph of §1.1. of op.cit. is the superposition of all the graphs $G(x)$, $x \in \Omega$.

***Theorem 4.*** Assume that (4) has several stationary states. Then there exists $x \in \Omega$ such that $G(x)$ contains a positive circuit.

**5. Discrete models**

In [16] Theorem 1 and [17] Theorem 2, it is shown that the stationary states of some piecewise linear models can be described by fixed points of a map $\Omega \to \Omega$, where $\Omega$ is a product of $n$ finite sets $\Omega_1, \ldots, \Omega_n$, with $\Omega_i$ of order $n_i + 1$, where $n_i$ is the number of thresholds values of the variable $x_i$. Theorem 4 gives therefore a proof that Thomas' rule holds for the fixed points of these discrete models.

For example, assume that $\Omega_i = \{0,1\}$ for all $i = 1, \ldots, n$ and let $F = (F_i) : \Omega \to \Omega$ be any map as in Theorem 1 above. Consider the system of piecewise-linear differential equations

(5) $$\frac{dx_i}{dt} = \tilde{F}_i(x) - x_i, \quad i = 1, \ldots, n,$$

where $\tilde{F}_i$ is defined below and $x$ lies in the open set $\tilde{\Omega} \subset \mathbf{R}^n$ consisting of those $(x_i)$ such that $x_i \neq 0$ for all $i$. Let

$$d : \tilde{\Omega} \to \Omega$$

be the map defined by

$$d(x)_i = (1 + \text{sign}(x_i))/2 .$$

We let

$$\tilde{F}_i(x) = \begin{cases} 1 & \text{if } F_i(d(x)) = 1 \\ -1 & \text{if } F_i(d(x)) = 0 \end{cases}$$

and $\tilde{F} = (\tilde{F}_i) : \tilde{\Omega} \to \mathbf{R}^n$. Notice that, for every $x \in \tilde{\Omega}$,

$$d(\tilde{F}(x)) = F(d(x)).$$

Assume that the system (5) satisfies the hypotheses of [16] §1, i.e., for every $i$, the function $\tilde{F}$ is written as a positive combination of sums and products of the functions $s_j(x), j = 1,\ldots,n$, where $s_j(x)$ is equal either to $s(x_j,0)$ or to $1 - s(x_j,0)$.

Then one can check that, for every $x \in \tilde{\Omega}$, if $y$ lies in the closure in $\mathbf{R}^n$ of the component of $\tilde{\Omega}$ containing $x$, the interaction graph $G(y)$ defined from $\tilde{F}$ as in §4 is contained in the interaction graph $G(d(x))$ of §1. Furthermore, according to [16] Theorem 1, if $d(x)$ is a fixed point of $F$ the point $x \in \tilde{\Omega}$ is an 'asymptotically stable' steady state of (5). From Theorem 4 we thus get a new (and quite indirect!) proof of Theorem 1, 2) under the above hypothesis. It would remain to decide which discrete models, and in particular which boolean models, can be obtained from the piecewise-linear models considered in [16] and [17].

## 6. Concluding remarks

The results above give support to the validity of Thomas' rule. They do not "prove" any kind of "biological law", since they depend on the quality of the models we used for describing gene networks, and these are, clearly, gross simplifications of the biological reality. For instance, we did not consider the role of chromatin conformation in the regulation of transcription. Neither did we include any discussion of the alternate splicing phenomenon. Therefore, it might be worth checking this rule in new and more refined setups. Let us mention a few possible extensions of the results presented here.

Gene networks do not appear in isolation. They are usually coupled with other biological networks, like the metabolic networks, and those involving interactions between proteins. People have tried to describe these mixed networks in a single picture (see for example [18] or [19]), but this is not so easy, since edges in these enlarged graphs do not have the same meaning as in the case of gene regulation. On the other hand, one should notice that an enlarged mixed network may well contain a positive circuit which is not visible in any smaller pure one, as soon as its vertices are of different nature.

Thomas' rule is of course valid for any system which can be modelled by one of the five methods described above. However, one has to be careful that the interaction graphs defined in these models by means of a mathematical recipe need not have an obvious intuitive meaning. In particular, they might not be simply related to the usual way of representing these systems (as noted in [20], Remark 5)), and a positive circuit needs not be visible in the traditional graphic representation. For instance, let us consider a chemical system containing, among others, two compounds $X$ and $Y$, and a reversible reaction

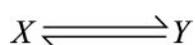

The concentrations $[X]$ and $[Y]$ obey the usual laws of chemical reactions, which look like

$d[X]/dt = a\ [Y] +$ *other terms*

$d[Y]/dt = b\ [X] +$ *other terms,*

where $a$ and $b$ are positive (products of a reaction rate with some concentrations). These equations are a special case of (1). If we compute the partial derivatives as in Section 2, we see that the corresponding interaction graph contains the motif which is a positive circuit. Thus, any reversible reaction gives rise to a positive circuit (although it does imply any kind of active chemical regulation). This type of situation has been considered in [21] and [22], where the authors gave examples of what they called "differentiation with no positive feedbacks". These are *not* counterexamples to Thomas' rule, since the corresponding interaction graphs do contain a positive circuit.

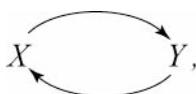

Another way to pursue our discussion would be to take into account the location of the proteins. Since the famous work of Turing in the fifties, many models of development have considered diffusion phenomena. It might be worth noting, though, that the scenario of spatial differentiation proposed by Turing and its followers requires that at least one of the chemicals is a self-activator. In that sense, there is still a positive circuit involved, and if we delete the diffusion terms in Turing's equations we get back to the situation described in Theorem 1. I do not know if diffusion in space can lead to differentiation in the absence of any positive circuit in the appropriate interaction graph.

A third direction is the following. All the models presented above are deterministic. Now, many recent works on gene expression insist on the importance of stochastic effects. These are manifest in the variation of expression levels from one cell to another. This forces us to view the binding of a protein to DNA, and the whole gene regulation, as a stochastic event. In [23], Gillespie proposed to represent the elementary chemical reactions in a gene network as a discrete jump Markov process, and, when there are enough copies of each protein, to approximate the chemical master equation by Fokker-Planck stochastic differential equations. It would be very interesting to decide if a variant of Thomas' rule still remains true in such a context. It is known that introducing stochasticity in a deterministic model allows for occasional switching between different stationary states (and such switches have been observed experimentally, see [24] for a survey). But can Gillespie's model lead to completely new scenarios of differentiation (violating Thomas' rule)?

Finally, one can also seek upper bounds for the number of possible stationary states in a gene network [11]. Such a bound is obtained in [25] when gene networks are modelled by means of IN and OR networks. Similar upper bounds cannot be valid for an equation like (1) since, obviously, the number of stationary states depends on the complexity of the function $F$ (e.g. its degree when $F$ is algebraic). For a general result on the complexity of gene networks, we refer to [26].

**Appendix A: Proof of Theorem 3**

We proceed by contradiction and assume that none of the graphs $G(z)$, $z \in \Omega$, contains a positive circuit. Then, according to [8] Lemma 2 i), all the principal minors of the Jacobian matrix of $(-F_i)$ at every point $z$ are nonnegative. Fix a constant $\varepsilon > 0$. By [8] (4) we conclude that all the principal minors of the Jacobian matrix of $(-F_i(x) + \varepsilon x_i)$ are positive in $\Omega$. By the univalence theorem of Gale-Nikaido ([27], [28] p. 20, a)), this implies that the map $(-F_i(x) + \varepsilon x_i)$ is a $P$-function on $\Omega$, i.e., when $x \neq y$ are two points in $\Omega$, there exists k $\in \{1,...,n\}$ such that

$$(x_k - y_k)(-F_k(x) + \varepsilon x_k + F_k(y) - \varepsilon y_k) > 0.$$

Since this is true for all $\varepsilon > 0$, there must exist k such that $x_k \neq y_k$ and

$$(x_k - y_k)(-F_k(x) + F_k(y)) \geq 0.$$

This means that $x_k \neq y_k$ and

$$(x_k - y_k)(F_k(x) - \gamma_k x_k - F_k(y) + \gamma_k y_k) \geq \gamma_k (x_k - y_k)^2,$$

hence

$$|F_k(x) - \gamma_k x_k - F_k(y) + \gamma_k y_k| \geq \gamma_k |x_k - y_k|.$$

This contradicts our hypothesis and proves Theorem 3.

**Appendix B: Proof of Theorem 4**

Following a suggestion of J.-L. Giavitto, we use an argument of approximation. For every $\theta \in \mathbf{R}$ and every integer $m \geq 0$ choose a differentiable function $s_m(x,\theta)$ on $\mathbf{R}$ such that $s_m(x,\theta) = s(x, \theta)$ when $|x - \theta| \geq \dfrac{1}{m}$, $s_m(x,\theta)$ is strictly increasing when $|x - \theta| < \dfrac{1}{m}$, and, given $x \in \mathbf{R}$ and $u \in s(x, \theta)$, for every $\varepsilon > 0$, there exists $m_0$ such that, if $m \geq m_0$, there is $\xi \in \mathbf{R}$ with

$$|\xi - x| < \varepsilon$$

and

$$|s_m(\xi, \theta) - u| < \varepsilon.$$

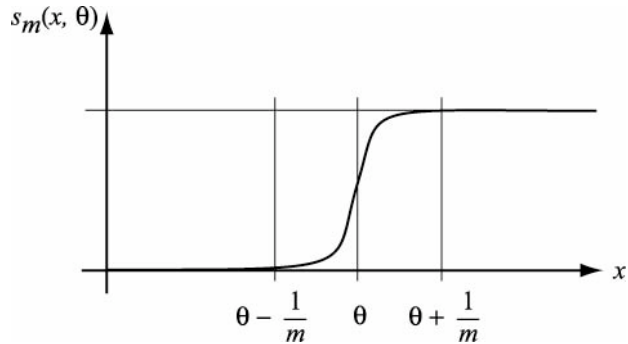

Assume $x \in \Omega$. It follows from (3) that, for every $u \in F(x)$ and every $\varepsilon > 0$, there exists $\xi \in \Omega$ such that, for all $i = 1, \ldots, n$,

$$|\xi_i - x_i| < \varepsilon$$

and

$$|F_i(\xi, m) - u_i| < \varepsilon,$$

where the functions $F_i(., m)$ are defined by the formula

$$F_i(\xi, m) = P_i(s_m(x_j, \theta_j^k)).$$

Assume now that the system (4) has two nondegenerate stationary states $x \neq y$ in $\Omega$. The assertion $0 \in (F_i(x) - \gamma_i x_i)$ means that $u_i \in F_i(x)$, where $u_i = \gamma_i x_i$. Similarly $v_i \in F_i(y)$ with $v_i = \gamma_i y_i$. For every $\varepsilon > 0$, and $m$ big enough, choose $\xi$ and $\eta$ in $\Omega$ such that, for all $i = 1, \ldots, n$,

$$|\xi_i - x_i| < \varepsilon, \qquad |\eta_i - y_i| < \varepsilon,$$

$$|F_i(\xi, m) - u_i| < \varepsilon, \quad \text{and} \quad |F_i(\eta, m) - v_i| < \varepsilon.$$

From these inequalities we conclude that

$$|F_i(\xi, m) - \gamma_i \xi_i - F_i(\eta, m) + \gamma_i \eta_i| < |u_i - \gamma_i \xi_i - v_i + \gamma_i \eta_i| + 2\varepsilon < 2\gamma_i \varepsilon + 2\varepsilon.$$

On the other hand, since $\gamma_i > 0$, when $x_i \neq y_i$ we can choose $\varepsilon$ small enough so that

$$2\gamma_i \varepsilon + 2\varepsilon < \gamma_i |\xi_i - \eta_i|.$$

It follows that $(F_i(., m))$ satisfies the assumption of Theorem 3, hence its interaction graph contains a positive circuit somewhere in phase space. It remains to notice that, for every $z \in \Omega$, when $m$ is big enough, the interaction graph of $(F_i(., m))$ at z is the same as the one of F at some point in $\Omega$.

# References

[1] Milo R., Shen-Orr S., Itzkovitz S., Kashtan N., Chklovskii D., Alon U., Network Motifs: Simple Building Blocks of Complex Networks, Science, Vol. 298, 2002, pp. 824-827.

[2] Delbrück M., Discussion, in Unités Biologiques douées de Continuité Génétique, Vol. 33, Editions CNRS, Lyon, 1949.

[3] Thomas R., On the relation between the logical structure of systems and their ability to generate multiple steady states or sustained oscillations, Springer Ser. Synergetics, Vol. 9, 1981, pp. 180-193.

[4] Thomas R., Kaufman M., Multistationarity, the basis of cell differentiation and memory. I. Structural conditions of multistationarity and other nontrivial behaviour, Chaos, Vol. 11, 2001, pp. 170-179.

[5] Snoussi E.H., Necessary conditions for multistationarity and stable periodicity, J. Biol. Syst., Vol. 6, 1998, pp. 3-9.

[6] Gouzé J.-L., Positive and negative circuits in dynamic systems, J. Biol. Syst., Vol. 6, 1998, pp. 11-15.

[7] Cinquin O., Demongeot J., Positive and negative feedback : Striking a balance between necessary antagonists. J. Theor. Biol., Vol. 216, 2002, pp. 229-241.

[8] Soulé C., Graphic requirements for multistationarity, ComplexUs, Vol. 1, 2003, pp. 123-133.

[9] Rémy E., Ruet P., Thieffry D., Graphic requirements for Multistability and Attractive Cycles in a Boolean Dynamical Framework, 2005, Preprint.

[10] de Jong H., Modeling and Simulation of Genetic Regulatory Systems: A Literature Review, Journal of Computational Biology, Vol. 9 (1), 2002, 67-103.

[11] Kaufmann S.A., The Origins of Order: Self-Organization and Selection in Evolution, Oxford University Press, 1993.

[12] Shih M.-H., Dong J.-L., A combinatorial analogue of the Jacobian problem in automata networks, Advances in Applied Math., Vol. 34 (1), 2005, pp. 30-46.

[13] Glass L., Classification of biological networks by their qualitative dynamics, J. Theor. Biol., Vol. 54, 1975, pp. 85-107.

[14] Gouzé J.-L., Sari T., A class of piecewise linear differential equations arising in biological models, Dynamical Systems, Vol. 17, 2002, pp. 299-316.

[15] De Jong H., Gouzé J.-L., Hernandez C., Page M., Sari T., Geiselmann H., Qualitative simulation of genetic regulatory networks using piecewise-linear models., Journal Math. Biol., Vol. 66, 2003, pp. 301-340.

[16] Snoussi E.H., Qualitative dynamics of piecewise-linear differential equations: a discrete mapping approach, Dynamics and Stability of Systems, Vol. 4, 1989, pp. 189-207.

[17] Snoussi E.H., Thomas R., Logical identification of all steady states: the concept of feedback loop characteristic states, Bulletin of Mathematical Biology, Vol. 55, 1993, pp. 973-991.

[18] Puchalka J., Kierzek A.M., Bridging the gap between stochastic and deterministic regimes in the kinetic simulations of the biochemical reaction networks, Biophysical Journal , Vol. 86, 2004, pp.1357-1372.

[19] Chaouiya C., Remy E., Thieffry D., Petri net modelling of biological regulatory networks, CMSB 2005 (Workshop, Computational Methods in Systems Biology), 3-5 avril, Edimbourg.

[20] Thomas R., The role of feedback circuits: positive feedback circuits are a necessary condition for positive real eigenvalues of the Jacobian matrix, Ber. Busenges. Phys. Chem., Vol. 98 (9), 1994, pp. 1148-1151.

[21] Markevich N.I., Hoek J.B., Khodolenko B.N., Signaling switches and bistability arising from multisite phosphorylation in protein kinase cascades, Journal of Cell Biology, Vol. 164 (3), 2004, 353-359.

[22] Karmakar R., Bose I., Graded and Binary Responses in Stochastic Gene Expression, arXiv q-bio. OT/0411012, 2004.

[23] Gillespie D.T., The chemical Langevin equation, J. Chem. Phys., Vol. 113 (1), 2000, pp. 297-306.

[24] Rao C.V., Wolf M.W., Arkin A.P., Control, exploitation and tolerance of intracellular noise, Nature, Vol. 420, 2002, pp. 231-237.

[25] Aracena J., Demongeot J., Goles E., Fixed points and maximal independent set on AND-OR networks. Discr. Appl. Math., Vol. 138, 2004, pp. 277-288.

[26] Grigoriev D., Vakulenko S., Complexity of gene circuits, Pfaffian functions and morphogenesis problem, Comptes-Rendus Acad. Sci., ser. 1, Vol. 337 (11), 2003, pp. 721-724.

[27] Gale D., Nikaido H., The Jacobian matrix and global univalence of mappings, Math.Ann., Vol. 159, 1965, pp. 81-93.

[28] Parthasarathy T., On global univalence theorems, Lecture Notes in Math., Vol. 977, Springer, 1983.